\documentclass{article}

\usepackage{PRIMEarxiv}

\usepackage[utf8]{inputenc} 
\usepackage[T1]{fontenc}    
\usepackage{hyperref}       
\usepackage{url}            
\usepackage{booktabs}       
\usepackage{amsfonts}       
\usepackage{nicefrac}       
\usepackage{microtype}      
\usepackage{lipsum}
\usepackage{fancyhdr}       
\usepackage{graphicx}       
\graphicspath{{media/}}     
\usepackage{amsmath}  
\usepackage{amssymb}  
\usepackage{amsfonts} 
\usepackage{hyperref} 
\title{A Design Trajectory Map of Human-AI Collaborative Reinforcement Learning Systems: Survey and Taxonomy
\thanks{\textit{\underline{Citation}}: 
\textbf{}} 
}

\author{
  Zhaoxing Li \\
  School of Electronics and Computer Science\\
  University of Southampton \\
  Southampton\\
  \texttt{\{zhaoxing.li\}@soton.ac.uk} \\
}

\begin{document}
\maketitle

\begin{abstract}
Driven by the algorithmic advancements in reinforcement learning and the increasing number of implementations of human-AI collaboration, Collaborative Reinforcement Learning (CRL) has been receiving growing attention. Despite this recent upsurge, this area is still rarely systematically studied. In this paper, we provide an extensive survey, investigating CRL methods based on both interactive reinforcement learning algorithms and human-AI collaborative frameworks that were proposed in the past decade. We elucidate and discuss via synergistic analysis methods both the growth of the field and the state-of-the-art; we conceptualise the existing frameworks from the perspectives of design patterns, collaborative levels, parties and capabilities, and review interactive methods and algorithmic models. Specifically, we create a new \textit{Human-AI CRL Design Trajectory Map}, as a systematic modelling tool for the selection of existing CRL frameworks, as well as a method of designing new CRL systems, and finally of improving future CRL designs. Furthermore, we elaborate \textit{generic Human-AI CRL challenges}, providing the research community with a guide towards novel research directions. The aim of this paper is to empower researchers with a systematic framework for the design of efficient and 'natural' human-AI collaborative methods, making it possible to work on maximised realisation of humans' and AI's potentials.
\end{abstract}


\section{Introduction}

With the rapid development of Artificial Intelligence (AI) in recent years, the mainstream media holds two opposing views: AI will 'save the world' \cite{skene2019artificial} or 'destroy' it \cite{yudkowsky2008artificial}. AI is described as the 'saviour', to free humans from labour, while it is also described as the 'devil' who takes away workers’ jobs \cite{zarkadakis2015our}. Regardless of one's point of view, AI is playing an increasingly significant part in the future world. \textit{Weak AI}, \textit{strong AI}, and \textit{super AI} are three stages of AI development, as proposed by John Searle \cite{sloman1986did}. Due to the limitations of current technology, Searle believes that we are supposed to have been and still be in the 'weak AI' stage for a long time. That is, at this current stage, AI often performs much worse than humans in highly complex decision-making tasks that require considerations of morality and risk, but much better in tasks with well-specified feedback and large scale data. Therefore, the two extreme situations described by the media are still far from the current stage that we've achieved \cite{skene2019artificial, yudkowsky2008artificial}. Exploring thus the way for humans and AI to better cooperate, with the goal of complementing each other's shortcomings, may provide the best way forward for the immediate future.

A common classification of Artificial Intelligence algorithms is supervised learning, unsupervised learning, and reinforcement learning \cite{gibney2016google}. Problems involving decision-making generally lie in the field of reinforcement learning \cite{sutton2018reinforcement}, and how humans and AI agents can cooperate and complement each other's shortcomings is particularly important. While the interaction between humans and AI agents is an emerging research direction, the research on the interaction between humans and computers has a long history. The community has proposed several patterns of human-computer interaction. For example, in 1983, Hollnagel and Woods proposed the Cognitive Systems Engineering (CES) model \cite{hollnagel1983cognitive}. In 1991, Schmidt \textit{et al.} created a conceptual paradigm classifying human-computer collaboration into three levels: augmentative, integrative, and debative \cite{schmidt1991cooperative}. In 2009, Johnson \textit{et al.} proposed a co-active design pattern in human-AI joint activity \cite{johnson2009joint}.

Over the last few years, collaborative or interactive reinforcement learning has become a new field within the machine learning regime. Some excellent recent survey studies on Collaborative Reinforcement Learning (Collaborative RL, or CRL) have emerged, demonstrating this new field's importance. They cover a wide range of issues including CRL in general such as \cite{amershi2014power}, and CRL applied in specific domains, such as safe RL \cite{garcia2015comprehensive}, inverse RL \cite{gao2012survey}, and explainable RL \cite{puiutta2020explainable}. Other studies concentrate on specific design methodologies such as user feedback and testbeds of the environment \cite{leike2018scalable}. When performing a brief search on Google Scholar with the keywords 'interactive AI', 'collaboration', 'reinforcement learning', and 'HCI' (Human-Computer Interaction) for the period from 2011 to 2022, we found that there are many surveys or literature reviews published. For example, only between January 2020 and October 2022, 5 surveys or literature reviews were published. Najar and Anis reviewed reinforcement learning based on human advice \cite{najar2020reinforcement}. Gomez and Randy's work focused on human-centered reinforcement learning \cite{li2019human}. Puiutta and Erick presented a review of explainable reinforcement learning \cite{puiutta2020explainable}. Arzate and Christian presented a survey on the design principles and open challenges of interactive reinforcement learning \cite{arzate2020survey}. Suran and Shweta consecrated on collective intelligence \cite{suran2020frameworks}. It can be observed that this research direction is gaining increasing attention from the community. However, surveying \textit{collaboration between} humans and AI agents is being overlooked, let alone identifying the probable future direction of the growth in this field. To bridge this gap, we thus aim to address the following research question:

\textbf{\emph{How may designers approach the construction process of human-AI collaborative reinforcement learning systems in an structured manner?}}

To answer this research question, we summarise existing collaboration approaches and offer our own perspectives and proposals. We look at classic human-machine interaction strategies that have had a significant effect on the evolution of the Human-Computer Interaction (HCI) field. We intend to give scholars and industry practitioners with a design toolkit that combines archetypes and specific tools in a micro-view \cite{classen2008s} (see Figure \ref{category}). Furthermore, our study introduces \textbf{Human-AI Collaborative Reinforcement Learning Design Trajectory Map}, (see Figure \ref{map}), a new categorisation approach and systematic modelling tool that seeks to suggest research objectives for \textbf{the next generation of Human-AI Collaborative Design}. Similar to how builders require the blueprint design as well as instructions on how to plan different functional parts and choose various types of materials for the house, the \emph{Design Trajectory Map} provides researchers with a comprehensive review regarding the \emph{design patterns} for Human-AI Collaborative Reinforcement Learning systems (see Section \ref{Design Pattern}) and guidance on how to customise the characteristics of different components to meet their specific requirements (see Sections \ref{Parties} and \ref{Capabilities}), as well as how to customise the algorithmic models (see Sections \ref{Algorithmic Models}) and the interactive methods (see Section \ref{Interactive Methods}).

\begin{figure}[h]
  \centering
  \includegraphics[width=8cm]{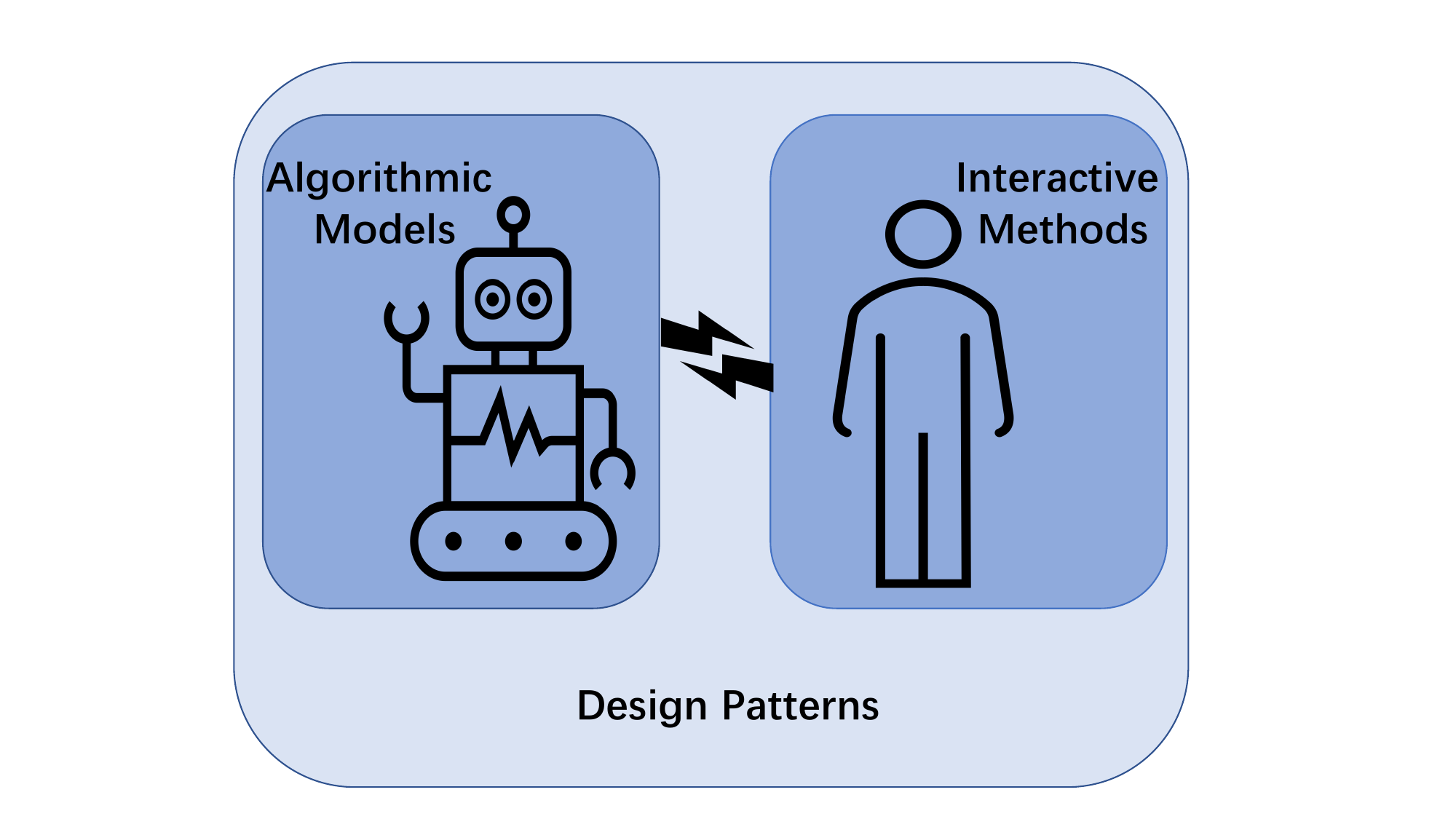}
  \caption{Human-AI collaboration Design Model: From a human perspective, we focus on how humans interact with AI agents; from an AI agent's perspective, we focus on how AI agents accept human instructions or suggestions in algorithm implementation; and from a collaboration pattern perspective, we focus on what kind of way that humans and AI collaborate.}
  \label{human-ai}
\end{figure}

This study builds on our previous work \cite{li2021survey} published at DIS '21: Designing Interactive Systems Conference 2021. In that work, we propose a Human-AI Design Model that designs a CRL model from three different perspectives: Human, AI agent, and Design pattern, which is a straightforward and effective method (See Figure \ref{human-ai}). In subsequent research, we found that in order to build a CRL system from macro to micro, we lacked the considerations of collaborative levels, parties and capabilities, which are crucial for designing the functions and details of each functional party. Therefore, in this work, we add a new Section \ref{Capabilities}. In addition, based on the original study, We improved in depth and topics covered. we create a more innovative comprehensive framework and a more complete taxonomy, covering from design patterns to algorithms, as well as increasing the review with 55 publications, accounting for 41 percent new literature. As a result, the primary contributions of this survey are as follows:
\begin{enumerate}
\item First, we summarise the most significant \textbf{Human-AI Collaborative Design Patterns}, which might help academics and practitioners in the HCI field.
\item Second, we present the \textbf{Collaborative Reinforcement Learning (CRL) Design Trajectory Map}, a \textit{novel CRL Classification and Taxonomy} as a systematic modelling tool, to assist researchers in selecting and improving new CRL designs.
\item Third, we take stock and summarise the most recent \textit{Collaborative Reinforcement Learning algorithms}, analysing the state-of-the-art at the start of this new decade. 
\item Fourth, as a \textit{roadmap to good Human-AI Collaboration}, we identify several general CRL problems for future study in this field.
\end{enumerate}

\section{Background}

Reinforcement Learning is derived from theories of \textit{animal learning} and \textit{parameter disturbance adaptive control} \cite{arzate2020survey}. The intuition is: if an agent's actions result in positive rewards (reinforcing signals), this type of behavior will be reinforced, increasing the agent's inclination to repeat the behavior in future acts. The goal of Reinforcement Learning is to train the agent to find the optimal strategy for each discrete state while maximizing the expected discounted rewards \cite{sutton2018reinforcement}. Mathematically, a reinforcement learning process can be described as a Markov Decision Process (MDP), defined by the tuple $\mathcal{M = \langle S, A, T, R, \gamma \rangle}$, which is a cyclical process, where an agent takes action $\mathcal{A}$ to change its state $\mathcal{S}$ to obtain a reward $\mathcal{R}$ from the process of interacting with the environment; $\gamma$ is the discount factor; $\mathcal{T}: S \times A \mapsto \operatorname{Pr}[S]$ is the transition function; the expected long-term reward follows policy $\pi$, represented as the Q-function $Q^{\pi}(s, a)$, which is computed as:

\begin{equation}
Q^{\pi}(s, a) = \mathbb{E}_{\pi} \left[ \sum_{t=0}^{\infty} \gamma^t R_t \mid s_0 = s, a_0 = a \right]
\end{equation}

$Q^{*}(s, a)=\max _{\pi} Q^{\pi}(s, a)$ is an optimal value function. Any optimal policy $\pi ^{*}$ that maximises the expected reward for each state is the solution to the MDP.

Reinforcement Learning differs from the other two types of Machine Learning: Supervised Learning and Unsupervised Learning. In \textbf{Supervised Learning}, a model learns the mapping relationship between input $X$ and label $y$ through a set of paired and labelled data, to solve \textit{Regression} and \textit{Classification} problems. In \textbf{Unsupervised Learning}, a model learns unlabelled data without any guidance, to solve \textit{Association} and \textit{Clustering} problems, for discovering underlying patterns of the data. In \textbf{Reinforcement Learning}, a model learns the mapping relationship between \textit{states} and \textit{actions} (non-predefined data), to solve \textit{Exploitation} or \textit{Exploration} problems. The mapping directs the model, or the agent, to make optimal decisions based on these \textit{states} towards \textit{maximising cumulative rewards} \cite{sutton2018reinforcement}. The learning process emphasises the interactions between the \textit{agent} and the \textit{environment} that gives `reward signals' during the agent's continual or exhaustive attempts of all the possible strategies to be adopted in a certain 'state', rather than directing the agent how to create the 'correct' action \cite{schwalbe2020survey}. A 'reward signal' is usually a \textit{scalar signal} and an assessment of the \textit{quality} of the generated action by the agent. This way, the \textit{agent} learns knowledge from the \textit{environment} through \textit{discrete feedback} of \textit{actions}, which it uses to optimise the parameters that might lead to an optimal result. With minimal information given by the external environment, the \textit{agent} must learn by its own interactions with the \textit{environment}, frequently from the ground up. If the 'reward signal' $r$ and the 'action' $A$ were known, these corresponding representation-label data might be utilised to train a model using supervised learning, however, it is often impracticable to exhaust all conceivable \textit{actions} in an \textit{environment} and create the corresponding '\textit{reward signals}'. This is where Reinforcement Learning may help. Reinforcement learning often beats \textit{Supervised Learning} in scenarios where the discrete action space is small,such as the game of Go or Atari \cite{adadi2018peeking}. 

Bellman proposed the mathematical theory of dynamic programming in 1955 \cite{bellman1955dynamic}. The Bellman condition was considered to be the crucial theoretical foundation for reinforcement learning. Then, in 1957, Bellman proposed Markov Decision Processes (MDPs) \cite{bellman1957markovian}, which are now used in most reinforcement learning algorithms. After the 1960s, the concepts of reinforcement and reinforcement learning gradually appeared in the literature. In 1963, a system called STeLLA was developed, which allows trial and error learning through interactions with the environment \cite{andreae1963stella}. Michie proposed an early reinforcement learning system called MENACE. In 1975, Holland proposed an adaptive system based on the selection principle in his book 'Adaptation in Natural and Artificial Systems' \cite{holland1992adaptation}. This is regarded as one of the most significant events in the evolution of reinforcement learning. The book also included genetic algorithms, which aided in the development of optimization algorithms.

Rummery and Niranjan suggested SARSA, or state-action-reward-state-action, in 1994 on the basis of Q-learning. In terms of decision-making, SARSA is similar to Q-learning, however it differs in terms of the updating method. SARSA employs an on-policy method, whereas Q-learning employs an off-policy one. \cite{rummery1994line}. Thrun \textit{et al.} introduced the Monte Carlo positioning method in 1999, which uses probability to solve the robot positioning problem \cite{dellaert1999monte}. Compared to traditional grid methods, it is more efficient and saves memory \cite{rummery1994line}.

With the advancement of computer power and the advancement of deep learning, numerous approaches combining deep learning with reinforcement learning have lately been presented. In 2013, Mnih \textit{et al.}, from the Deep Mind team, proposed Deep Q-Learning (DQL) \cite{mnih2013playing}. This approach employs Q-learning to discover the appropriate control rules after transferring data from high-dimensional sensory input to a convolutional neural network to extract features. This team's AlphaGo defeated the world Go champion with a score of 3:0 in 2017. In December of same year, the more advanced Alpha Zero achieved AlphaGo master by self-learning without the assistance of human knowledge in only 21 days, and exceeded all versions after 40 days \cite{silver2017mastering}. Since then, \textit{Reinforcement Learning} has made great progress \cite{gibney2016google}. In 2021, Chen \textit{et al.} proposed a method that transforms reinforcement learning into a sequence modelling problem called the decision transformer model \cite{chen2021decision}. This has gradually been applied in fields such as games \cite{lample2017playing}, robotics \cite{kober2013reinforcement}, computer vision \cite{bernstein2018reinforcement}, natural language processing (NLP) \cite{branwen2020gpt}, and recommender systems \cite{rohde2018recogym}. For example, the Open AI$\footnote{OpenAI website:\ www.openai.com}$ team created an interactive reinforcement learning method that used human feedback, to learn summarisation \cite{stiennon2020learning}. Another recent project proposed by this team, GPT-3, has also made revolutionary achievements in the field of NLP \cite{branwen2020gpt}.

Due to its strong potential and firm theoretical foundation, Reinforcement Learning has recently been one of the research areas in AI technologies that attract the most attention \cite{sutton2018reinforcement}. However, it faces many challenges. Currently, on the one hand, Reinforcement Learning only works well when the environment is definite, i.e., the state of the environment is fully observable. In particular, there are defined rules in games like Go, and the action space is discrete and constrained. In other words, the agent needs a great degree of prior knowledge, to understand its state in a complex environment \cite{zhang2021survey}. On the other hand, even if the agent has been given well-specified feedback, the inexplicability and incomprehensibility caused by the agent's unconscious is still inadequate for the agent to decide on the precise next action \cite{arzate2020survey}. Furthermore, most applications of Reinforcement Learning to date have only been for playing games, such as chess and Atari.

\section{Methodology and Scope}

\subsection{Literature Collection}

This review focuses on hitherto undiscovered areas of research on collaborations between humans and AI agents. We further refine our data pool, specifically narrowing the selection to the following selected target areas: \textit{HCI}, \textit{Human-AI Collaboration}, \textit{Reinforcement Learning}, and \textit{Explainable AI}, published in the recent decade - the time period between 2011 and 2022 from Google Scholar. In total, our search yield 237 articles using keywords including \textit{collaborative reinforcement learning}, \textit{interactive reinforcement learning}, \textit{human-computer interaction}, and \textit{design patterns}. They were published in journals and conferences, including top venues such as TOCHI$\footnote{The website of TOCHI: https://dl.acm.org/journal/tochi}$, IJHCS$\footnote{The website of IJHCS: http://dblp.uni-trier.de/db/journals/ijmms/}$, AAAI$\footnote{The website of AAAI: www.aaai.org/}$, CHI$\footnote{The website of CHI: chi2021.acm.org/}$, UbiComp$\footnote{The website of UbiComp: www.ubicomp.org/}$, UIST$\footnote{The website of UIST: uist.acm.org/}$, and IEEE$\footnote{The website of IEEE:www.ieee.org/}$. Following a manual review of the title and abstract of each article, we eliminated 103 articles as irrelevant, leaving 134 articles as the source of this survey.

\subsection{Human-AI Collaborative Reinforcement Learning Classification}

We use an \textit{inductive method} to organise the literature we collected, and proposed our new classification method, inspired by the traditional human-machine interaction research. Previous work by Najar and Anis mainly targeted physical interaction between humans and machines from a human perspective \cite{najar2020reinforcement}. 
In the early stage of computer and engineering, there was no involved concept of AI. So, there is only research on human-machine interactions. Cruz \textit{et al.} proposed that the human-AI interaction is a kind of human-machine interaction in general \cite{cruz2016multi}. In this paper, we have collected paradigms of human-machine interaction in the early stage, which also could apply to human-AI interaction. Therefore, in the following section, we will use the concept of human-AI interaction in a unified manner. In the work of Arzate and Chirstian, more attention was paid to the algorithmic model of the AI agent \cite{arzate2020survey}. After reviewing the literature of human-machine interaction in traditional engineering, we found that the interaction between humans and AI corresponds to these design patterns. In particular, Schmidt's model \cite{schmidt1991cooperative} not only combines the interactive methods and algorithmic models, but also provides different design ideas, according to different human-AI collaborative levels. Based on the common characteristics of the classification of these literature, we derive the new \textit{Human-AI Collaborative Reinforcement Learning (CRL) Classification} (see Table \ref{tab:inter}).

\begin{table*}
  \caption{\textit{CRL Classification} applied to the Pool of Papers about Collaborative RL, published between 2011-2022}
  \label{tab:inter}
  \begin{tabular}{ccl}
    \toprule
    & Collaborative RL  &References\\
    \midrule
    & Cognitive Systems Engineering Patterns& \cite{hollnagel1983cognitive}\\
     Design  & Bosch and Bronkhorst's Patterns &  \cite{van2018human}\\
    Patterns& Coactive Design Patterns& \cite{johnson2014coactive} \\
    & Schmidt' s Patterns& \cite{zieba2010principles} \\
    
     \midrule
     Collaborative& Augmentative Level collaboration& \cite{moon2003intelligent},   \cite{adadi2018peeking}, \cite{wu2020regional}, \cite{verma2018programmatically},\cite{liu2018toward},\cite{madumal2019explainable}\\
     Levels  & Integrative Level collaboration & \cite{johnson2014coactive},  \cite{zieba2010principles}, \cite{van2019six}, \cite{chacon2019pandemic}, \cite{cardona2014games}, \cite{liang2019implicit}, \cite{schmidt1991cooperative} \\
     and& Debative Level collaboration &  \cite{schmidt1991cooperative},
     \cite{kling1981routine}, \cite{irving2018ai},  \cite{garcia2015comprehensive}, \cite{frederiksen1999dynamic},  \cite{arzate2020survey},  \cite{arumugam2019deep}\\
     Parties& Collaborative types&  \cite{hagl2020safe}, \cite{dafoe2020open}\\
     
     \midrule
     
    & Understanding & \cite{hovi1998games},\cite{albrecht2018autonomous}, \cite{ng2000algorithms}, \cite{armstrong2017occam}, \cite{amin2017repeated},\cite{ bai2016information},\cite{de2017negotiating},\cite{bard2020hanabi} \\
    Collaborative & Communication & \cite{dafoe2020open},\cite{wagner2003progress},\cite{sukhbaatar2016learning},\cite{foerster2016learning},\cite{cao2018emergent},\cite{zhang2021survey} \\
     Capabilities & Commitments & \cite{fearon1995rationalist},\cite{sen1985goals}\cite{north1993institutions},\cite{fearon1995rationalist},\cite{bagwell1995commitment},
      \cite{jackson2011reasons}, \cite{criswell2014kcofi}, \cite{luu2016making}, \cite{frantz2016institutions} \\
      & Institutions & \cite{smith1980contract},\cite{horling2004survey},\cite{ferber1999multi},\cite{bond2014readings},\cite{vazquez2005organizing},\cite{dutting2015payment},\cite{dutting2019optimal},\cite{tacchetti2019neural},\cite{amir2015multi} \\
     
    \midrule
    Interactive & Explicit Methods & \cite{knox2009interactively}, \cite{loftin2016learning}, \cite{macglashan2017interactive}, \cite{thomaz2005real},\cite{thomaz2008teachable},\cite{knox2009interactively},\cite{warnell2018deep}, \cite{cruz2016multi}\\
     Methods  & Implicit Methods & \cite{voyles1999gesture}, \cite{moon2003intelligent}, \cite{li2020facial}, \cite{gadanho2003learning}, \cite{arakawa2018dqn}, \cite{gordon2016affective}, \cite{goyal2019using}, \cite{maclin1996creating}, \cite{kuhlmann2004guiding},  \cite{williams2018learning}
    \\
   & Multi-model Methods& \cite{quek2002multimodal}, \cite{weber2018real},\cite{isbell2001social}, \cite{lessel2019enable},\cite{ho2015teaching}\\
    
     \midrule
     & Reward-based methods& \cite{thomaz2006reinforcement},
     \cite{knox2009interactively},  \cite{argall2009survey},\cite{arakawa2018dqn}, \cite{gordon2016affective},  \cite{goyal2019using}, \cite{maclin1996creating}\\
    Algorithmic & Policy-based Methods &  \cite{griffith2013policy}, \cite{macglashan2017interactive},
     \cite{arumugam2019deep}, \cite{gordon2016affective}, \cite{goyal2019using}\\
    Models & Value Function based methods& \cite{taylor2011integrating}, \cite{li2018introspective}, \cite{mnih2013playing}, \cite{silver2017mastering}, \cite{moon2003intelligent} \\
    & Exploration-process methods& \cite{thomaz2008teachable},  \cite{thomaz2006adding}, \cite{maclin1996creating}, \cite{kuhlmann2004guiding}, \cite{williams2018learning}, \cite{johnson2014coactive}, \cite{schmidt1991cooperative} \\
  
  \bottomrule
\end{tabular}
\end{table*}

\subsection{Human-AI Collaborative Reinforcement Learning Taxonomy}

\begin{figure}[h]
  \centering
  \includegraphics[width=\linewidth]{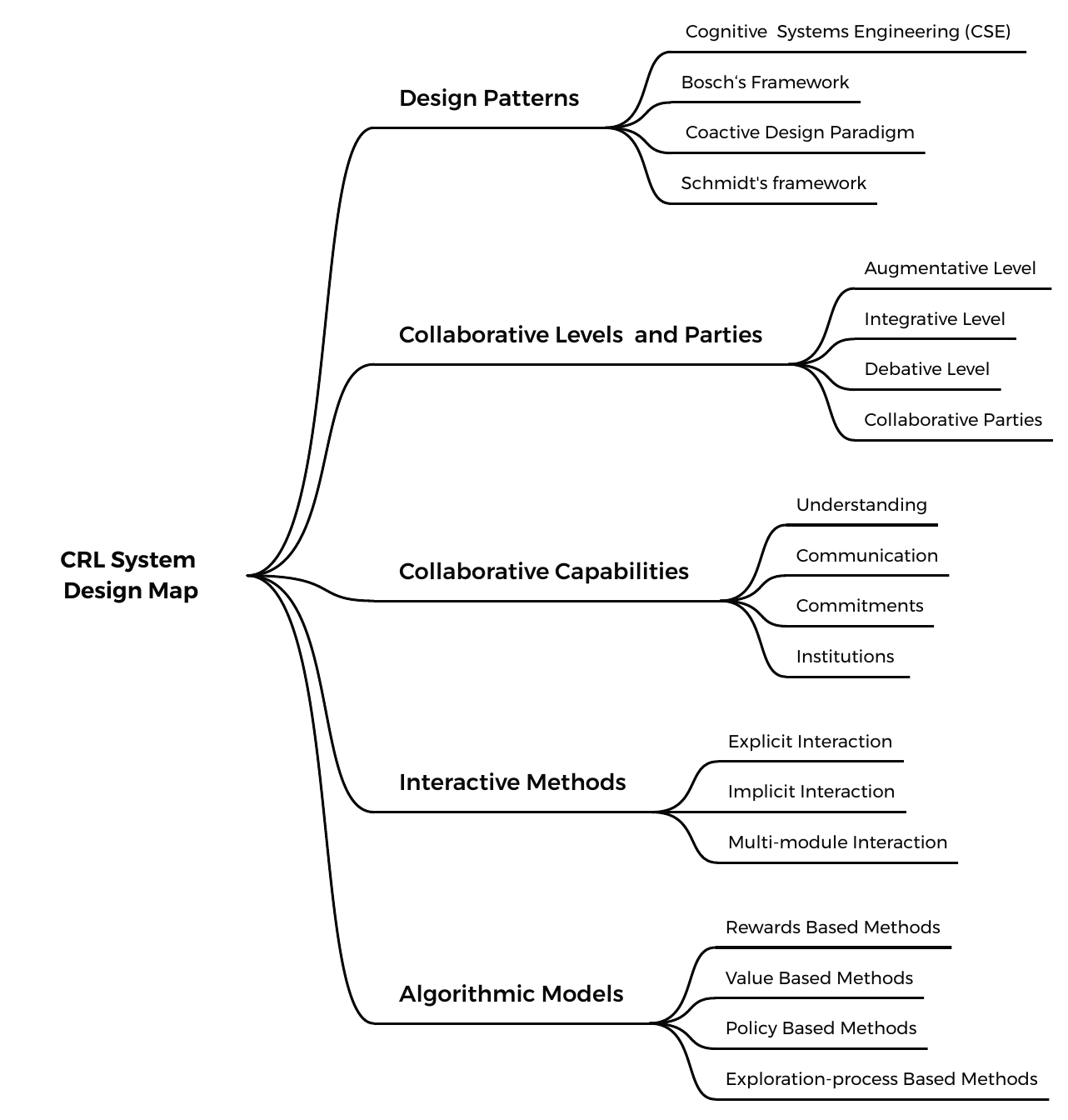} 
  \caption{A new \textit{CRL taxonomy} for interactive methods and design patterns}
  \label{category}
\end{figure}

We incorporate past work in a novel way to create a new taxonomy. We draw on Schmidt's Machine Interaction Pattern \cite{schmidt1991cooperative}, Dafoe's Collaboration Parties Classification Model \cite{dafoe2020open}, and Arzate's Algorithmic Classification Model Collaboration Method \cite{arzate2020survey}, to generate a novel taxonomy method from coarse to fine granularity. Based on this approach and populate them with representative works from the literature, for a structured approach (see Table \ref{tab:inter}), we define five axes: include \textit{Design Patterns}, \textit{Collaborative Levels and Parties}, \textit{Collaborative Capabilities}, \textit{Interactive Methods}, and \textit{Algorithmic Models}. These five axes are then used to create a taxonomy, as shown in Figure \ref{category}, which might be used as a systematic modelling tool for HCI researchers and practitioners to select and improve their new CRL designs.

\section{Human-AI Collaborative Design Patterns} \label{Design Pattern}
Human-AI collaborative design patterns may be used to provide an efficient and repeatable approach for building human-AI collaborative systems \cite{ferenc2005design}. Reliable design patterns might increase these systems' quality, reusability, and maintainability. In this article, we collect the most recognised design patterns in the literature of \textit{human-machine collaboration} for academics and practitioners to populate the CRL Taxonomy as shown in Figure \ref{category}.

In comparison to human-AI collaboration, human-machine collaboration has long been a source of concern for researchers. In the early stages of the development of human-machine interactions, domain experts believed that the collaboration between humans and machines was a physical, lower-level type of collaboration \cite{schmidt1991cooperative}. Specifically, machines were used by humans exclusively through physical contact, in the lack of feedback between machines and humans, which might be viewed as a kind of \emph{unidirectional interaction}. After decades of development, several recognised human-machine collaboration design patterns have been created, which we summarise below.

\subsection{CSE Pattern} Cognitive Systems Engineering (CSE), coined by Hollnagel and Woods, acts at the level of cognitive functions \cite{hollnagel1983cognitive}. CSE is the first framework proposed to analyse the human-machine information exchange interaction. CSE is a framework for human-machine collaboration, where machines 'plan and explore' using the knowledge or information provided by humans. This engineering method suggests that human-machine collaboration occurs at a conscious level of communication. It is a perceptual mode in which the machine is employed as a sensory extension to assist with human activities.

At this level, the major challenge is to identify appropriate interactive methods to optimise human information processing. However, CSE has been constrained in that it has only explored basic and low-level communications, leaving out more complicated problems and environments.

\subsection{Bosch and Bronkhorst's Pattern} Bosch and Bronkhorst defined three levels of Human-AI collaboration: 1) \emph{unidirectional interaction}, in which humans assist machines or machines explain themselves to humans; 2) \emph{bi-directional interaction}, and 3) \emph{collaboration} between humans and machines \cite{van2018human}. The vast majority of the currently existing methods have only addressed the first level.

This framework's contribution is to provide a viewpoint on the directions of collaboration between humans and machines. Furthermore, it constructs the direction based on the roles that humans and machines play in a task, with one being the subject of a task and the other aiding the opposing side. It is believed to help in the development of more efficient communication methods. For example, if it is a human-centred framework, more considerations should be given to how to transform a 'machine language' into interpretable information such that humans can better understand; whereas if it is a machine-centred framework, more considerations should be given to what human knowledge could improve machine efficiency.

\subsection{Coactive Design Pattern} Johnson \textit{et al.} proposed a Coactive design pattern in human-AI joint activities. They experimented with a collaboration system from the perspective of observability, predictability, and directability \cite{johnson2014coactive}. \emph{Observability} concerns the ability of both robots (or AI agents) and humans to observe each other's pertinent aspects of status, as well as the knowledge of the team, tasks, and the environment. \emph{Predictability} refers to the state that the actions of both robots (or AI agents) and humans can be predicted such that they may rely on each other's actions to perform their own actions. \emph{Directability} refers to the ability of both robots (or AI agents) and humans to direct each other's behaviour in a complimentary manner.

This framework is similar to the Bosch and Bronkhorst's framework in that it considers the direction of interactions and divides it into different levels. However, it is limited due to a lack of robustness and security considerations.

\subsection{Schmidt's Pattern} Schmidt believes that collaboration should be tailored to diverse needs, fulfil different functions, and be carried out in a variety of ways depending on the circumstances. collaboration may be summarised as follows: 1) the augmentative level, in which one role in the partnership (human or AI agent) assists the other in performing tasks; 2) the integrative level, in which both sides of the team share information and assist each other in completing tasks together; and 3) the debative level, in which tasks are completed through debate and negotiation between humans and AI agents, especially when dealing with complex issues \cite{zieba2010principles}.

This framework considers not only the information exchange direction in the collaboration, but also different levels of collaboration, as well as the robustness, security, and potential ethical considerations.

\section{Collaborative Levels and Parties} \label{Parties}
\begin{figure}[h]
  \centering
  \includegraphics[width=9cm]{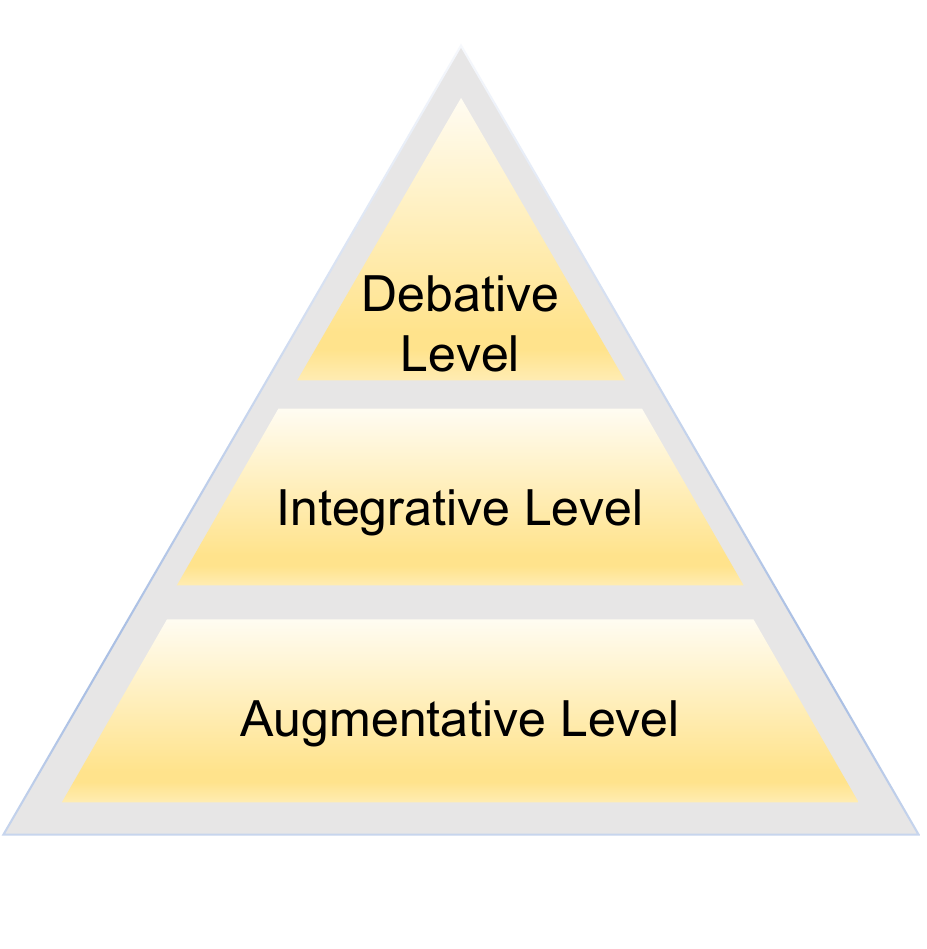}
  \caption{Triangle of different collaborative levels: the first level is Augmentative Level collaboration; the second level is Integrative Level collaboration; and the third level is Debative Level collaboration.}
  \label{Different levels}
\end{figure}

The patterns described above frame the modes of human-AI collaboration from different perspectives. They are also very comparable in terms of compartmentalising collaboration modes or methods, i.e., into single-direction assistance, bi-directional collaboration, and higher-stage fused collaboration. In this survey, we utilise a fusion viewpoint that combines interactive methods and design patterns based on Schmidt's collaboration pattern to classify the current collaborative reinforcement learning techniques. Schmidt’s model is divided into three levels: augmentative, integrative, and debative. We drew a pyramid model based on Schmidt's model (see Figure \ref{Different levels}). We highlight significant research that have emerged at each level and the issues that should be examined in the first three sub-sections that follow. We also discuss the characteristics, advantages, and disadvantages of these different methods, as well as how to develop new methods in the future.

Apart from the classification of collaborative levels where humans and AI agents are both viewed as a whole, we also discuss collaboration from a \emph{micro perspective} based on the framework proposed by Dafoe \textit{et al.}, where diverse constellations of humans and AI agents are discussed, which we refer to as “collaborative parties” in the final sub-section.

\subsection{Augmentative Level collaboration}
collaboration at the \emph{Augmentative Level} entails one partner compensating for the shortcomings of the other \cite{schmidt1991cooperative}. AI has showed considerable promise in large-scale data processing with well-defined rules, as well as in natural-perception fields such as digital image recognition \cite{bernstein2018reinforcement}, natural language processing \cite{wang2018deep}, among others. Nevertheless, in a complex ambiguous environment, AI performance lags considerably below that of humans. The \emph{Augmentative Level} approaches offered by the community are mostly divided into two types. First, AI takes the lead in decision-making, while humans assist AI in enhancing processing efficiency. In this case, humans use prior knowledge to help the agents specify the state space and efficiently obtain rewards from the complex environment. Second, humans play the primary role in decision-making, with AI assisting in the process. In this case, the AI agents explain the tactics used to help humans make faster decisions in a simple environment. At the sub-level of \emph{humans helping AI agents} improve efficiency, we will categorise how they communicate based on which parts of the algorithm humans' help can be injected. At the sub-level of \emph{AI agents helping humans}, we mainly focus on how AI agents may inform humans about why they make particular decisions.

\subsubsection{Human->AI}
The most essential aspect in the role of humans supporting AI agents in decision-making has been how to efficiently deliver information to AI agents while reducing human weariness to a minimal. Up to this point, many human-AI collaborative reinforcement learning algorithms have been proposed, which may be categorised into \textit{explicit interaction modal}, \textit{implicit interaction modal}, and \textit{multi-modal methods} based on different forms of interactions (detailed in Section \ref{Design Pattern}). Finding a better way for humans to directly interact with AI agents is still an essential research priority.

\subsubsection{AI->Human}

In the task of AI agents assisting humans in decision-making, the most challenging problem lies in interpretability. Interpretability (or explainability) refers to the degree to which humans can understand the rationale underpinning machines' decision-making \cite{holzinger2018machine,li2021survey}. The interpretability of AI models refers to the clarification of the internal mechanism and the understanding of the results. The more interpretable the model, the easier it is for people to trust it \cite{puiutta2020explainable,li2022simstu}. Its significance may be seen in the following aspects: in the \textit{modelling phase}, interpretability may assist developers in understanding the learning process, comparing alternative algorithms, optimising the procedure, and fine-tuning the models; in the \textit{operation phase}, AI agents can explain the internal mechanism and interpret the model outcomes to the decision-maker (i.e., humans). Consider a decision-making recommendation model, before the model runs, multiple interpretable algorithms with their respective advantages can be provided to humans to choose from; and after the model is trained, the model must explain to humans why it recommended a specific solution given a specific context.

Patterns underlying the above problems lie under the umbrella of eXplainable AI (XAI), which is commonly regarded as critical for the practical deployment of AI models. DARPA launched the XAI in 2016 \cite{adadi2018peeking,li2023broader,li2023deep}. The basic objective of XAI is to create machine learning models that, when combined with proper explanation techniques, will allow humans to better comprehend and eventually accept and trust the model's predictions. The literature generally proposes two types of explainability: 1) \textit{transparent models}, which are embedded inside the operation of the AI algorithms, leading to explainability by design, applied to simpler AI algorithms with less accurate results; and 2) \textit{post-hoc models}, which are performed after initial models have been trained. This type of methods is usually more efficient, but it is less reliable than \textit{transparent models} \cite{wu2020regional,li2023sim,li2023towards}.

At present, there are a few intrinsic interpretability Reinforcement Learning methods. Verma \textit{et al.} introduced a Programmatically Interpretable Reinforcement Learning method (PIRL) \cite{verma2018programmatically,li2024lbkt}. This method is an upgrade of traditional Deep Reinforcement Learning (DRL). In DRL, due to the 'black-box' nature of neural networks, it is difficult to represent policies. To tackle the 'black-box' challenge, PIRL introduces an advanced human-readable programming language to define neural network policies. Shu \textit{et al.} introduced a hierarchical and interpretable multi-task reinforcement learning framework, where a complex task is broken into several sub-tasks and then a hierarchical strategy is used to complete the learning with 'weak supervision' from humans. By breaking a task into sub-tasks and thus making a learned strategy traceable to them, and by explaining the relationship between different hierarchies of the sub-tasks, this method builds intrinsic interpretability.

Compared with intrinsically intepretable Reinforcement Learning methods, post-hoc methods are simpler in algorithm structure and more efficient in the computing process. At present, many post-hoc methods have been proposed. For example, Liu \textit{et al.} proposed an explainable DRL method based on linear model U-trees \cite{liu2018toward}. This is a stochastic gradient descent framework for explaining complex models by using linear model U-trees to fit Q-functions. There is also a Soft Decision Tree (SDT) method, which provides post-hoc explanations by extracting policies. Madumal \textit{et al.} introduced an explainable method through a causal lens. In this framework, an AI agent learns to play StarCraft II, a large dynamic space strategy game \cite{madumal2019explainable,wang2020humans}. To generate an explanation, they simplify the entire game states to four basic actions and nine basic states, and then use these basic causal factors to construct an explanation for why the AI agent chooses action A over action B.

\subsection{Integrative Level collaboration}
Integrative collaboration entails using the various advantages of both parties to complete a task. At this level, humans and AI agents are regarded as being interdependent. The main task is broken into several sub-tasks, and humans and AI agents can perform just those that they are skilled at \cite{schmidt1991cooperative}. At the integrative level, humans and AI agents play equal roles in the system. Information exchange at this level is generally referred to as 'communication' in the literature \cite{schmidt1991cooperative,wang2023exploring}.

In the following sub-sections, first, we summarise the communication methods in this cooperative pattern. Then, we discuss how to make the communicating parties trust each other. On this basis, the system needs resilience to enhance its robustness in order to better deal with the complex conditions in the real world.

\subsubsection{Communication}
A grand challenge of collaborative reinforcement learning is how humans and AI agents communicate with each other. Only when communication is seamless can they make decisions on the next actions following each other’s feedback. Liang \textit{et al.} proposed an implicit human-AI collaboration framework based on Gricean conversational theory \cite{wilson1981grice} to play the game Hanabi. The AI agent must cooperate with the human to win the game. In this framework, the AI agent tries to understand the implied meaning of human's natural language suggestions in a dialogue box \cite{liang2019implicit}.

Cordona-Rivera and Young proposed an AI Planning-based Gameplay Discourse Generation framework to achieve communication between human players and the game \cite{cardona2014games}. Pablo and Markus proposed an approach of Human-AI collaboration by planning and recognition of the plan \cite{chacon2019pandemic}. Johnson \textit{et al.} proposed a testbed for joint activities. The unique feature of this testbed is that it can be applied not only in interactive experiments for multiple agents but also in interactive experiments between humans and agents \cite{johnson2009joint}. A series of works were carried out on this testbed to study the collaboration of humans and agents in a team. For example, Matthew \textit{et al.} introduced the relationship between the interdependence and autonomy in a human-AI collaboration system \cite{van2019six}.

\subsubsection{Trust}
Based on the established communications, how to make the partners trust each other to complete the task is also crucial. Although the community has not yet proposed a clear definition of trust between humans and AI agents, it is generally regarded as a psychological state \cite{schmidt1991cooperative}. Johnson \textit{et al.} proposed a Coactive Design framework for human-AI joint activities. In their framework, the authors proposed a collaboration system following the perspective of observability, predictability, and directability \cite{johnson2014coactive}. These components are critical for humans and AI agents to collaborate in a trustworthy manner.

\subsubsection{Resilience}
Resilience is another essential feature in human-AI collaboration. On the premise of communication and mutual trust, in complex problems, with possible delays and information noise, how to establish a resilient mechanism to make the system more robust is crucial. An effective human-machine collaboration mechanism should be able to diagnose a problem quickly and provide remedial explanations after the problem occurs so that the system can get back on course \cite{johnson2014coactive}.
Zieba \textit{et al.} proposed a mechanism to measure the resilience of human-machine systems, that is, the ability to anticipate, avoid, and recover from accidents to a normal state \cite{zieba2010principles}. This is instructive to design a cooperative system, as it is necessary to consider how the system responds to emergencies and thus recovers quickly.

\subsection{Debative Level collaboration}
Debative models come into play when humans and AI agents hold different opinions on decision-making in a task, and they debate to find the optimum solution based on their differing knowledge and understandings. Models are often required to meet the following requirements. First, humans and AI agents share a unified goal, and achieving that goal is the primary task. For both parties, a debate without a unified goal is meaningless. Second, both parties have strong justifications for their decisions and have insights into a problem based on their respective cognitive models. Third, both parties can effectively communicate and explain their decisions to each other. Communication and interpretability are the premises of the debate. Fourth, there are clear evaluation criteria to measure the outcome from a debate to ensure an optimal result. Fifth, both parties can learn and adjust their own knowledge after a debate to achieve better results in the future \cite{schmidt1991cooperative,wang2023user}.

As knowledge-based decisions are fragile and controversial, it is necessary to debate the results \cite{kling1981routine}. In a complex and uncertain environment, a full debate will better demonstrate the advantages and disadvantages of different decisions. collaboration at the level of debate requires that both humans and AI agents have sufficiently high professionalism in a specific complex domain. Reinforcement learning algorithms based on this level are scarcely studied in the literature, but we expect that as the field progresses, this form of collaboration will attract more attention.

Geoffrey \textit{et al.} introduced a framework that enables two agents to debate with each other, with a human judge deciding who to trust in the end \cite{irving2018ai}. Although it has not yet been applied to the debate between humans and agents, this framework meets the requirements outlined above. In their experiment, the two agents attempted to persuade human judges to believe their judgments on the MNIST data \cite{deng2012mnist}. First, the goal of the two agents were unified. Second, the two agents have different judgments based on their own algorithmic perceptions. Third, both agents are able to generate simple explanations to persuade human judges. Fourth, human judges have intuitive knowledge to make accurate judgments. This experiment is enlightening for future research, especially in human-agent and multi-agent debate collaboration.

\subsection{Collaborative Parties}
\begin{figure}[h]
  \centering
  \includegraphics[width=9cm]{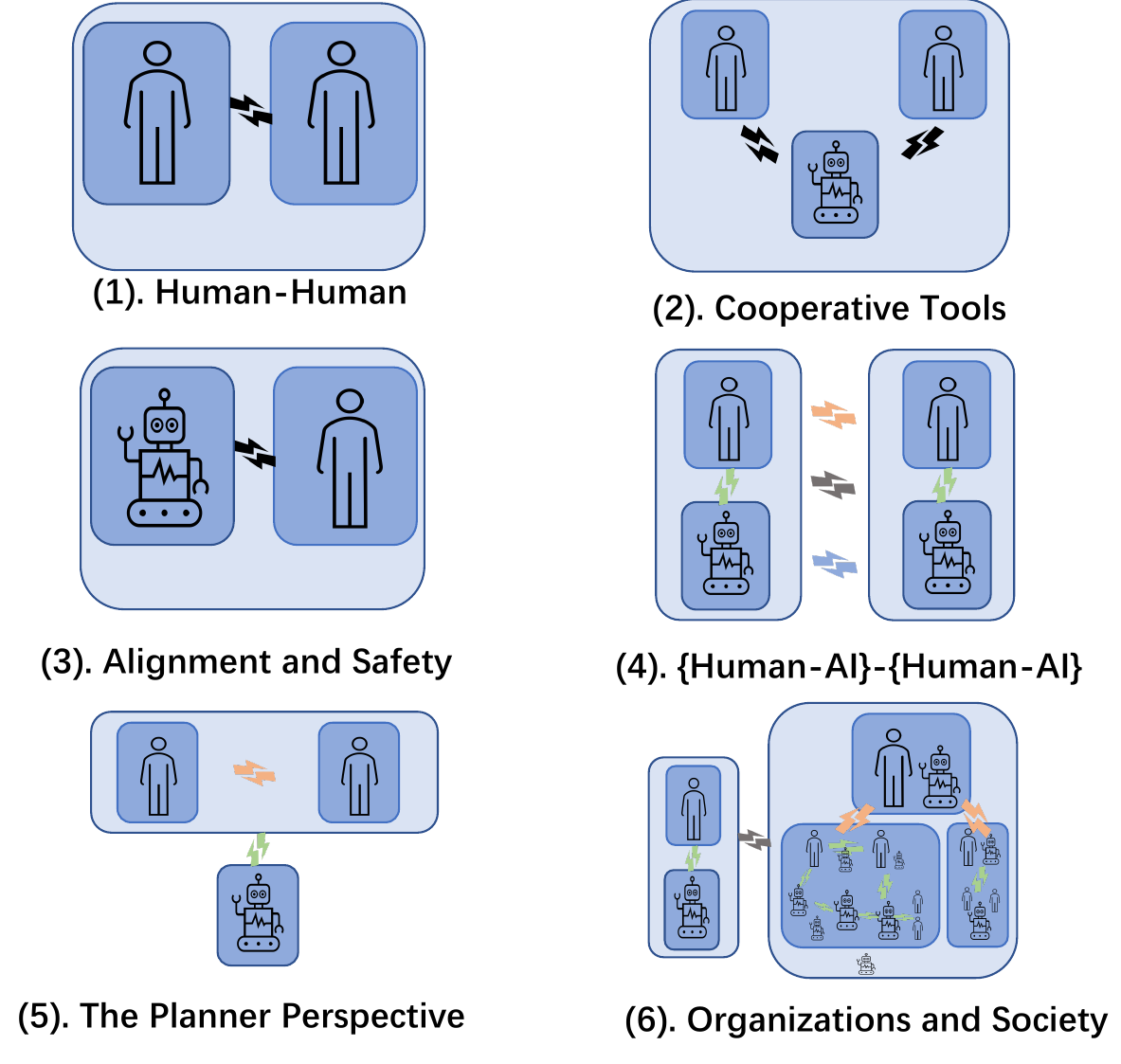}
  \caption{Different parties in the process of collaboration.}
  \label{Different paties}
\end{figure}

The different levels of human-AI collaboration take a macro view of human and AI, looking at them both as a whole. The collaboration of humans with AI, on the other hand, can be split down into different combinations of parties from a micro perspective or in considerations of practical scenarios. For example, future scenarios could include interactions between a human and several AI-agents, interactions between human groups and AI-agent groups, or more diversified fusion of the two. Therefore, in this section, we will discuss the types of interactions between humans and AI agents from the micro perspective (AI agents-agents, human-AI agents, human-human, and more complex constellations). Dafoe \textit{et al.} categorises cooperative roles into three categories: AI-agent, Human and Organisations. Collaborative types based on the number of roles involved into the collaboration into six types \cite{dafoe2020open} (see Figure \ref{Different paties}):

\begin{enumerate}
    \item Human-Human collaboration: the classic human-to-human collaborative model; 

    \item Cooperative Tools: the AI agent is used to enhance collaboration, such as language translation.

    \item Alignment and Safety: the AI agent acts like an assistant to help humans solve problems, such as the relationship between vehicles and humans in autonomous driving.

    \item {Human-AI}-{Human-AI} collaboration: With the development of 5G technology and AI technology, large-scale human groups and AI groups may cooperate in the foreseeable future, such as the level 5 autonomous driving \cite{hagl2020safe}.

    \item The Planner Perspective: This approach is to strengthen the collaboration and infrastructure of the entire society from the planner perspective of social construction, rather than the collaboration between an individual and a single AI, e.g., social media and network communications.

    \item Organisations and Society: Collaboration could have a more complicated structure, with multiple types of hierarchical collaboration or a complex internal structure.

\end{enumerate}

\section{Collaborative Capabilities} \label{Capabilities}
In the previous section, we discussed the different levels of human-AI collaboration from a macro perspective, where both human and AI agent are viewed as integral parts. However, in the real scenarios or from an micro perspective, there may be interactions between a human and several AI-agents, or interactions between human groups and AI groups. Therefore, in this section, we will discuss the types of human-AI interactions from the micro perspective (i.e., AI agents-AI agents, human-AI agents, human-human, and more complex constellations), as well as discuss what kinds of collaboration capabilities the agent require for groups interactions.

Dafoe \textit{et al.} divides collaboration capabilities of agents into 4 types: Understanding, Communication, Commitments, and Institutions \cite{dafoe2020open}.

\subsection{Understanding} 
In human-AI Collaboration, an AI agent's ability to understand the environment and predict the consequences of its actions is crucial for reaching mutually beneficial results. In game theory, there are many discussions about how important it is to understand multi-role collaboration. For example, in Nash equilibrium, each strategy is required to be the best response after fully understanding other's strategies \cite{hovi1998games}. Moreover, under the constraint of partial information, Bayesian Nash Equilibrium and Perfect Bayesian Equilibrium provide a solution for how multi-role collaboration should enable other participants to better understand each other's strategies \cite{barrett2020deciding}.

In collaborative reinforcement learning, the most important type of understanding is learning the preference of the other AI agents, namely the values, goals, reward functions of the other parties. Humans understand how to better provide feedback or rewards to the AI agent in order to help it converge faster and function more efficiency. Some AI researchers attempted to directly learn the AI agent's behaviour. For example, Albrecht's study \cite{albrecht2018autonomous} summarised how humans observe the AI agent's behaviour in order to understand the agent. Inverse Reinforcement Learning. or IRL, is a type of research in which AI agents are oblivious or indifferent to humans \cite{ng2000algorithms}. This type of method requires humans to inject their prior knowledge \cite{armstrong2017occam} or control the AI agent to make a few first steps \cite{amin2017repeated, bai2016information}.

Besides, in a more complex decision-making environment, humans may consciously or unconsciously hide their opinions or ideas causing significant challenges for AI agents. There have been some studies on the application of human recursive mind-reading methods in negotiations to overcome this challenge \cite{de2017negotiating}, and some studies have applied this method to the game Hanabi to improve the collaboration between humans and AI agents \cite{bard2020hanabi}.

\subsection{Communication}
Understanding and collaboration could be difficult to achieve without effective communication. AI agents may often get a better understanding of others' behaviour, intentions, and preferences by communicating directly with them rather than just observing and interacting with them on a regular basis. The finding of Pareto-optimal equilibrium may be made easier as a result of information exchange \cite{dafoe2020open}.

\paragraph{Common Ground}
A common ground is necessary for collaboration. The message sender and receiver should use the same communication protocol so that each may understand the meaning of the other's message.

Many studies have been conducted on machine-to-machine  communication problems, which are usually referred to as emergency communication \cite{wagner2003progress, sukhbaatar2016learning, foerster2016learning,cao2018emergent,wang2024comparative}. However, there are few studies on how to establish common ground and effective communication between humans and AI agents. The establishment of a common ground is arguably the most difficult challenge \cite{dafoe2020open}.

\paragraph{Bandwidth and Latency}
The bandwidth of communication refers to the volume of data that may be transferred in a given duration of time \cite{dafoe2020open}. Latency refers to the time it takes for a message to be transmitted and received \cite{dafoe2020open}. How to enhance bandwidth and minimise latency in human-AI collaboration has long been studied, and some promising techniques have been proposed, including brain-computer interface technologies, which are designed to connect human brain directly via hardware to maximum bandwidth with the shortest possible latency \cite{zhang2021survey}.

\subsection{Commitments}
The aforementioned capabilities of Understanding and Communication strive to overcome the difficulties in collaboration caused by inaccurate or inadequate information. Collaboration, even with abundant information, may still fail. Social scientists have identified "commitment issues", or the inability to make credible threats or promises, as a primary cause of collaboration failure. Prominent research even claims that the problem of commitment is the most significant impediment to rational AI agent collaboration \cite{fearon1995rationalist}. A substantial volume of research has explored the commitment issues that affect collaboration \cite{sen1985goals,north1993institutions,fearon1995rationalist,bagwell1995commitment,jackson2011reasons}.

Many different studies tried to build commitments between Human and AI. Some studies have attempted to develop a commitment contract method between humans and AI agents based on semantics \cite{criswell2014kcofi, luu2016making, frantz2016institutions,wang2024comparative}.

\subsection{Institutions}
Obtaining the requisite understanding, communication, and commitment for collaboration often necessitates the use of a social framework. In economics and politics, this social framework is typically referred to abstractly as \textit{institutions} \cite{dafoe2020open}.

\paragraph{Decentralised Institutions}
In decentralised institutions, there is no single institutions centre, and each individual is connected to each other, and continuously encourages the construction of the structure through the interaction with each other. Additionally, many multi-agent system construction methods have been proposed to aid in multi-agent systems communication, planning, and decision-making \cite{smith1980contract, horling2004survey,ferber1999multi,bond2014readings,vazquez2005organizing,wang2024impact}.

\paragraph{Centralised Institutions}
Centralised institutions involve a centralised authority that can define the rules and limit other participants \cite{dafoe2020open}. The multi-agent systems research community attempts to build a collaborative mechanism amongst agents using approaches based on centralised institutions \cite{rosenschein1994rules}. Several studies investigated the use of centralised multi-agent systems in automatic auction systems \cite{dutting2015payment,dutting2019optimal, tacchetti2019neural}. The method of centralised multi-agent path-finding technique could be utilised in autonomous vehicle obstacle detection in the future \cite{amir2015multi}.

\section{Interactive Methods} \label{Interactive Methods}
Traditional reinforcement learning methods require excessive training-time in complex environments, and their applications are often confined to scenarios with clear rules. An effective way to mitigate these limits is the use of different strengths of human and AI and complementing one other's inadequacies. This approach is known as Collaborative Reinforcement Learning (CRL). CRL employs human-in-the-loop training to improve the performance of algorithms or to help humans improve decision-making efficiency \cite{arzate2020survey}. Recent CRL research has focused on developing AI that can communicate with humans in a more natural way \cite{arzate2020survey}. There are two types of interactive methods: explicit and implicit. In an explicit method, humans provide the AI agents with clear numerical feedback, explicitly. This method is preferable for AI agents since it allow them to process the feedback more easily, but it is likely to cause human fatigue due to the ambiguity of numerical representations, resulting in inefficiency in a long-term training process. In an implicit method, humans give feedback to AI agents through natural interactions such as posture and gaze, as opposed to explicit methods, which provide clear numerical feedback. This method places more demands on the AI agent, but it may improve the fatigue resistance of human trainers, allowing for long-term and stable collaboration \cite{isbell2001social}. Based on these unsolved problems, in this section, we present human-AI interactions from the perspective of interactive methods.

\subsection{Explicit Interactive Methods}
Currently, most AI agents learn from human feedback via explicit interactive methods. Humans provide feedback directly to the AI agent via keyboard, slider bar, or mouse to provide clear alpha-numerical feedback \cite{knox2009interactively,loftin2016learning,macglashan2017interactive,thomaz2005real}. For example, Thomaz and Breazeal proposed a method of sending feedback to the AI agent by using the mouse to click on the sliding bar \cite{thomaz2008teachable}. Knox and Stone proposed the TAMER framework, which allows an AI agent to learn from MDP and human advice by having a human trainer click the mouse to indicate the desired actions \cite{knox2009interactively}. These methods are more efficient than traditional reinforcement learning, and can achieve specific goals in complex environments with the assistance of humans.

However, the reaction time of human trainers may cause delayed feedback, leaving the AI agent unsure of which actions the human feedback was aimed at, especially for AI agents with frequent actions. A common solution is to set a delay parameter to express the past time-steps. For example, Warnell \textit{et al.} proposed a method to obtain the delay distributions of the human trainers to improve algorithm efficiency \cite{warnell2018deep}. Knox and Stone provided another way for estimating the delay: using a probability density function \cite{knox2012learning}. Moreover, these methods may be unfavourable to non-professional human trainers, who need to spend a significant amount of time learning the user interface and the meaning of feedback represented by each operation. Simultaneously, this kind of interactions can easily make human trainers impatient.

Human trainers can also provide explicit feedback to AI agents using hardware delivery methods \cite{cruz2016multi}, where feedback is be generally converted into a numeric value directly via the hardware devices, such as keyboards. However, a more user-friendlier method is for the AI agent to learn implicit feedback from natural interactions with trainers.

\subsection{Implicit Interactive Methods}
Aside from receiving feedback directly from human trainers via explicit interactive methods, the AI agents can also learn via implicit interactive methods.

Implicit interaction methods reduce the learning cost of human trainers, as they can directly participate in training the AI agents without specific learning. At the same time, a more natural way of interaction may reduce the fatigue of human trainers. Many implicit interactive methods have lately been proposed. For example, feedback can be based on natural language, facial expressions, emotions, gestures, and actions, as well as the incorporation of multiple natural interactive methods. In an ideal scenario, humans could train the AI agent in the same way that they interact with humans in the real world. Below, we summarise some of the most prominent implicit interactive methods.

\paragraph{Gestural Feedback.} Gestures are sometimes considered to be a form of unconscious human communication. It is also considered to be an effective way to complement other communication forms, and it is even more useful than other communication methods for users who are speech- or hearing-impaired. For example, Voyles and Khosla proposed a framework which can train robots by imitating human gesture \cite{voyles1999gesture}. Moon \textit{et al.} introduced a method of using gestures to command the AI agent to learn to control a wheelchair \cite{moon2003intelligent}. These methods are very friendly to human trainers and do not require any particular training on their part.

\paragraph{Facial Feedback.} Li \textit{et al.} trained a mapping model to map implicit emotions to various types of explicit feedback data. Facial expressions were marked with different types of feedback in advance, such as 1 for “happy” and 0 or -1 for “sadness” \cite{li2020facial}. Based on this work, Gadanho introduced a facial feedback reinforcement learning method based on an emotion recognition system. The system can learn to decide when to change or reinforce its behaviour with Q-learning by identifying human emotions \cite{gadanho2003learning}. Arakawa \textit{et al.} introduced the DQN-TAMER model, where an AI agent may obtain facial expressions via a camera, and then use the facial expression data to map different emotions as implicit rewards to improve learning efficiency \cite{arakawa2018dqn}. Veeriah \textit{et al.} proposed a method where the agent may analyse human facial features from camera images to gain additional rewards. As a result, the AI agent can quickly adapt to the user's facial changes in order to complete the task \cite{gordon2016affective}. One of the limitations of this kind of methods is that human emotions cannot be identified merely based on facial expressions, and there may be a delay in converting machine recognition expressions into feedback.

\paragraph{Natural Language Feedback.} When compared to facial expression and gesture tracking feedback methods, natural language feedback makes it easier to convert the token vector of the sentence into quantitative feedback. Natural language feedback can be transformed and applied to several aspects of reinforcement learning, such as rewards, values, and policies. Goyal \textit{et al.} introduced the LEARN (LanguagE-Action Reward Network) method, which is a reward shaping method \cite{goyal2019using}. In the state-action space of the task, if most of the reward signals are 0s, we call it the sparsity of rewards. Sparse rewards may cause the algorithm to converge slowly. AI agents need to interact with the environment several times and learn from a large number of samples to reach an optimal solution. One solution to this problem is to provide the AI agent with a bonus reward in addition to the reward function whenever the AI agent takes a right step toward the goal. This process is called reward shaping. Maclin and Shavlik proposed RATLE (Reinforcement and Advice, Consulting Learning Environment) \cite{maclin1996creating}, where the AI agent can translate human natural language suggestions into feedback for the Q-value function to accelerate the learning process. Kuhlmann proposed a method based on transforming natural language suggestions into an algorithm-understandable formal language to optimise the learning policy \cite{kuhlmann2004guiding}. In addition to the methods described above for transforming into different parts of the algorithm, natural language can also be used to directly guide the AI agent's learning policy. For example, Williams \textit{et al.} proposed an object-oriented Markov Decision Process (MDP) framework which can map the natural language to rewards feedback \cite{williams2018learning}.

\subsection{Multi-modal Feedback} 

The research above is focused on a single input interaction method. Multi-modal interactions, on the other hand, are more prevalent and efficient in day-to-day human-human interactions. Multi-mode communication has the following benefits. First, when a single-mode piece of information is disrupted by noise or occlusion, other modes can be used as information supplements. Second, when multi-modal interaction is available, it has the potential to improve the robustness and reliability of communication. Quek \textit{et al.} introduced a framework for analysing language's mutual support and accompanying gestures \cite{quek2002multimodal}. Cruz \textit{et al.} proposed a dynamic multi-modal audiovisual interaction framework that would allow humans to provide feedback using their voices and gestures \cite{weber2018real}. Griffith \textit{et al.} \cite{griffith2013policy} introduced a multi-modal interaction method based on hand gestures and speech recognition system, which was restricted to operating geometric objects on maps. Weber \textit{et al.} \cite{weber2018real} developed a dynamic audiovisual integration method that allows humans to input information via natural language and gestures. In the above experiments, multi-mode interactions generally outperformed single-mode interactions. Most of the current multi-mode interactions are merely a combination of two modes, such as any two of voice, gesture, sound, and vision. One of the problems of the above multi-modal methods is their inability to combine various forms of human feedback. The ability of humans to directly interact with AI agents using multiple methods at the same time remains unexplored. In the future, these multi-mode interactive methods can be combined in more forms to develop effective human-AI collaboration for a wider range of scenarios.

Some studies take into account the effect of human fatigue caused by increasing training time on the quantity and quality of feedback. As training duration grows, human trainers become exhausted, reducing the amount of feedback while simultaneously lowering the quality of the feedback \cite{isbell2001social, ho2015teaching}. Methods for encouraging human trainers to raise interaction excitement through gamification were proposed; such methods have been found to decrease weariness and effectively improve human trainers' efficiency \cite{lessel2019enable}.

\section{Algorithmic Models} \label{Algorithmic Models}

In the previous section, we analysed how humans provide feedback to AI agents. In this section, we categorise algorithmic models based on how agents receive and process human feedback.

\subsection{Reward-based Methods}
Reward-based methods accelerate the learning process by adjusting the reward that the AI agent receives from the environment. Concretely, after the AI agent receives feedback from the environment, humans can scale up or down the rewards based on their knowledge, potentially accelerating the learning process \cite{argall2009survey}. Computationally, the reward from human $H\left(s, a, s^{\prime}\right)$, is added to the reward from the environmental reward $R\left(s, a, s^{\prime}\right)$ to get the new reward $\bar{R}\left(s, a, s^{\prime}\right)$.
\begin{center}
\begin{equation}
\bar{R}\left(s, a, s^{\prime}\right)=R\left(s, a, s^{\prime}\right)+H\left(s, a, s^{\prime}\right),
\end{equation}
\end{center}

Thomaz and Breazeal proposed a method for non-expert human trainers to influence the AI agent's next action by providing a positive or a negative numerical reward. If the agent received a negative feedback, it would attempt to reverse the previous action in order to get a higher score \cite{thomaz2006reinforcement}. 

Knox and Stone first introduced the TAMER algorithm, which uses human demonstration as input to guide the AI  to perform better \cite{knox2009interactively}. Based on the TAMER method, Riku \textit{et al.} introduced a framework which combines deep learning method and TAMER, named DQN-TAMER, where rewards were shaped by the human's numerical binary feedback and environments \cite{argall2009survey}. Additionally, Arakawa \textit{et al.} investigated a facial expression function based on reward-shaped method, which is applied in a maze-like environment game \cite{arzate2021interactive}. The human trainers' facial expressions could provide feedback to the AI agents. The major shortcoming is that the recognition of the human facial expression is imprecise and intermittent.

Rosenfeld \textit{et al.} develops a heuristic function method, where the AI agent receive feedback generated by hand-engineered data from the human trainer \cite{rosenfeld2018leveraging}. The experiment's findings \cite{bianchi2013heuristically} indicate that heuristic functions may be a natural method for AI agents to learn from human trainers. The primary disadvantage of this approach is that it requires human trainers with extensive professional backgrounds and programming skills. It will be extremely hostile to non-professional users.

Reward-based methods can efficiently expedite the learning process in an environment with sparse rewards, but there are certain drawbacks, listed below. The first problem is "credit allocation", which is especially problematic in a rapidly changing environment where humans may be too slow to provide timely feedback. Therefore, the method's limitation remains how to map human rewards to corresponding actions. The second problem is "reward hacking", where the AI agent may achieve the greatest rewards by using ways that humans would not expect \cite{arzate2020survey}.

\subsection{Inverse Reward Design Methods}

The agent is constantly attempting to optimise the human-designed reward function. When designing the AI agent, human developers always set the reward function based on the experimental environment, but the AI agent always encounters a new environment. Using the original reward function designed by humans in a new environment may lead to poor convergence. Mindermann \textit{et al.} presented an inverse reward function in response to this issue \cite{mindermann2018active}. To obtain the true target, this method is based on the designed reward function and the trained MDP. This allows agents to effectively adapt to the new environment, eliminating the issue of reward hacking. More specifically, this method takes the designed reward function, the test environment model, and the MDP in the new environment as input. Then a Bayesian function maps the proxy rewards to the real rewards. The experiment in \cite{hadfield2017inverse} demonstrates that the inverse reward method could successfully boost the AI agent's learning efficiency.

\subsection{Policy-based Methods}
Policy-based methods modify the learning policy of the AI agent action process to encourage the action to fit what the human trainers expect \cite{arzate2020survey}. Human trainers may be aware of a large number of potential optimal actions $A$ in a given state $S$, the probability of human providing feedback to the AI agent can be denoted as $C$, where $0 < C < 1$. The difference between positive and negative human feedback can be expressed as $\Delta_{s, a}$. The probability that humans give policy feedback $\operatorname{Pr}_{c}(a)$ in a given state $S$ can be expressed as
\begin{center}
\begin{equation}
\operatorname{Pr}_{c}(a)=\frac{C^{\Delta_{s, a}}}{C^{\Delta_{s, a}}+(1-C)^{\Delta_{s, a}}}
\end{equation}
\end{center}

At present, the method that uses human critique for state and action pairs as input to shape agent policy is widely accepted. Griffith \textit{et al.} proposed an optimal policy method based on human feedback, a Bayesian method that takes as input critiques for each state and action pair \cite{griffith2013policy}. The experiments in \cite{loftin2016learning} suggest that this policy-based method outperforms other reward-based methods.

Krening and Feigh conducted an experiment in which they compared two different policy-based methods that could bring a better user experience \cite{krening2018interaction}. The first one is critique feedback method proposed by Griffith \cite{griffith2013policy}, and the second is their Newtonian action advice method \cite{krening2018interaction}. The result is that the method of action advice is better and the time required is reduced.

MacGlashan \textit{et al.} proposed a Convergent Actor-Critic method, COACH (Corrective Advice Communicated by Humans). This framework allows non-experts to use numerical binary feedback to formulate policies through corrective suggestions \cite{macglashan2017interactive}. Dilip \textit{et al.} proposed a deep COACH method based on the original COACH, which uses raw pixels as input to train the AI agent's policy. The authors argued that the use of highly representative inputs facilitates the application of the algorithm in more complex environments \cite{arumugam2019deep}.

When compared to reward-based methods, the advantage of policy-based methods is that they do not require specific feedback from humans to AI agents. Nevertheless, humans must determine which strategy may be the most effective assisting the AI agent. This may have higher requirements for the prior knowledge of human trainers.

\subsection{Value Function based Methods}
Value function based methods estimate future rewards to obtain the highest potential reward at the end of the task, by using human knowledge \cite{arzate2020survey}. They combine the value representing human preference with the value obtained by the AI agent from the environment to promote the learning process. Matthew \textit{et al.} proposed a method that combines human preference and agent value called Human-Agent Transfer (HAT) \cite{taylor2011integrating}. The algorithm generates a strategy based on recorded human trainer preferences, which it then applies to shape the Q-value function. This shaping process provides a stable reward for the state-action pair, in the Q-learning process. Brys \textit{et al.} proposed a method that uses human demonstrations as input for a value named RLfD. This method generates a Gaussian function by human demonstration to guide the exploration process of the Q($\lambda$) algorithm \cite{li2018introspective}.

Despite the fact that value function based methods are likely to be an effective way for minimising human feedback, there are now just a few studies based on it.

\subsection{Exploration Process based Methods}

Reinforcement learning is a method in which an AI agent needs to continuously interact with the environment and complete tasks based on rewards. This means that the AI agent needs to perform actions that it has never tried before. This process is referred to as the exploration process. In exploration process based methods, humans can increase the efficiency by reducing AI agent errors and unnecessary attempts \cite{argall2009survey}. Exploration process based methods aim to minimise the action space by injecting prior human knowledge to guide the AI agent’s exploration in order to increase learning efficiency. 

Thomaz and Breazeal conducted an experiment in the game Sophie's Kitchen to evaluate human guidance that helps the AI agent minimise its action space in order to enhance learning efficiency \cite{thomaz2008teachable}. The results suggest that employing human prior knowledge to limit low utility efforts is more efficient than using scalar reward functions \cite{thomaz2006adding}. Suay \textit{et al.} developed an upgrading approach in which the user may help exploration by highlighting goal states in the environment \cite{suay2011effect}. Yu \textit{et al.} proposed an approach, termed as action biasing, which leverages user feedback to stimulate the AI agent's exploration process. The sum of the agent and user value functions is employed as a value function, to incorporate human feedback into the AI agent's learning process \cite{yu2018learning}. These methods are considered effective, but they generally need to be trained by humans, and this training process requires a lot of professional knowledge and participation.

In general, collaborative reinforcement learning has shown great potential in improving the efficiency of decision-making tasks. However, further research is needed to determine how to build the environment models in which humans interact with AI agents. These models should consider not only the effectiveness and efficiency of interactive methods, but also interpretability, accountability, and possible ethical issues in decision-making. Therefore, in the following sections, we refer to the literature on the pattern of human-machine relations in the engineering field, and propose guidance for future development of collaborative reinforcement learning methods.

\section{Design Trajectory Map} 

\begin{figure}[h]
  \centering
  \includegraphics[width=\linewidth]{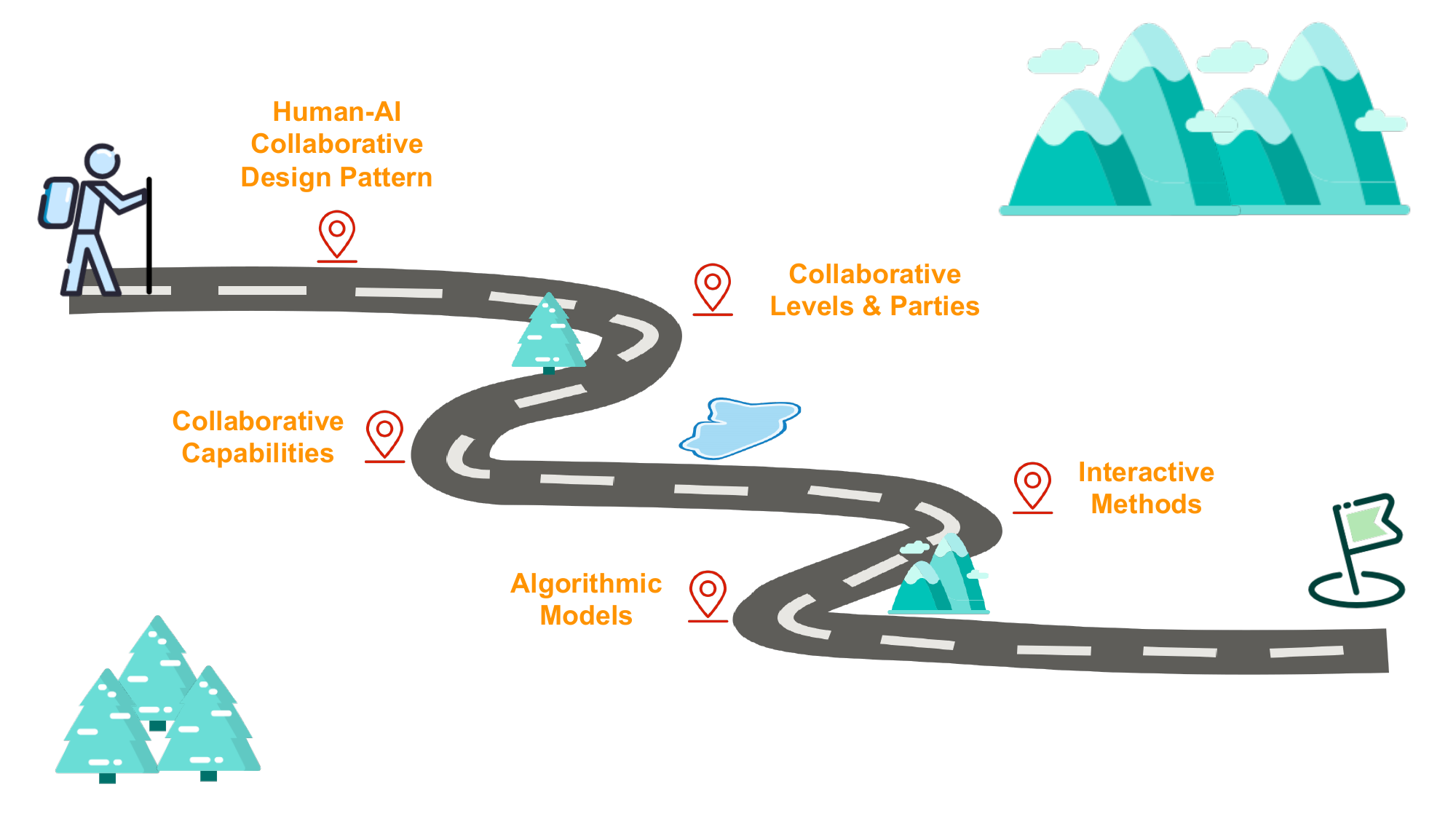}
  \caption{A Design Trajectory Map of Collaborative Reinforcement Learning Systems}
  \label{map}
\end{figure}

Based on the previous CRL taxonomy, we propose a novel \textit{CRL Trajectory Design Map} to guide researchers design CRL systems. When researchers start designing a human-AI collaborative reinforcement learning system, they could follow our \textit{CRL Trajectory Design Map} (Figure \ref{map}) step by step. First, they start with selecting a collaborative pattern from a macro perspective in the \textit{Design Patterns} (Section \ref{Design Pattern}) category. Next, they choose different collaborative levels and a number of the participants in the \textit{Collaborative Levels and Parties} (Section \ref{Parties}). After that, they choose the cooperative capabilities that every party should have in the \textit{Collaboration Capabilities} (Section \ref{Capabilities}). Finally, they select suitable interactive methods and algorithmic models for the specific task requirement categories of \textit{Interactive Methods} (Section \ref{Interactive Methods}) and \textit{Algorithmic Models} (Section \ref{Algorithmic Models}).

Figure \ref{category} presents our newly proposed CRL taxonomy, which contains the most commonly used and highly cited methods and design patterns in the CRL research area, and which can also be used as a Trajectory Map (see Figure \ref{map}) of designing collaborative reinforcement learning systems, as follows. Researchers may utilise our Trajectory Map to develop their architecture as they go from the top Design Patterns to the next, until the most detailed Algorithmic Models is selected. In the Map, the first part suggests \textbf{Design Patterns}, which are the most popular structure of human-AI collaborative frameworks in the CRL domain. These include cognitive systems engineering (CSE) \cite{hollnagel1983cognitive}, Bosch's framework \cite{van2018human}, the Coactive design pattern \cite{johnson2014coactive} and Schmidt's framework \cite{zieba2010principles}. The second part are the \textbf{Collaborative Levels and Parties}, while the Third part covers \textbf{Collaborative Capabilities}, which include understanding, communication, commitments, and institutions. The forth part is \textbf{Interactive Methods}, including explicit and implicit interaction methods, as well as multi-module interaction modes \cite{arzate2020survey}. The last part reflects \textbf{Algorithmic Models}, which contains reward-based methods \cite{argall2009survey}, value-based methods \cite{taylor2011integrating}, policy-based methods \cite{griffith2013policy}, and exploration-process-based methods \cite{thomaz2008teachable}. This taxonomy could be used as a systematic modelling tool for researchers and practitioners to select and improve their new CRL designs. They could choose an archetype in \textit{Design Patterns} for the overall architecture at start. Then, they could select a \textit{Collaborative Level and the numbers of the Parties} in the collaboration. After that, they could select the \textit{Collaborative Capabilities} that the AI agents should have, and select suitable \textit{Interactive Methods} and \textit{Algorithmic Models} that can meet the requirements of specific tasks. If researchers wish to learn about the most advanced technology developed in the last decade, they could check the classification we provide in Table \ref{tab:inter}.

\section{Future Work Recommendations}\label{Recommendations}
Reinforcement learning seems to have reached a plateau after experiencing a rapid development. It is difficult to improve the efficiency of AI agents in a complex environment without clear feedback. The research community has proposed some collaborative methods to overcome these obstacles. For example, humans deliver feedback to AI agents through hardware or sensors to improve algorithms efficiency; AI agents provide humans with explanations of decisions to improve the credibility of algorithms. However, research in this area is only at the beginning stage, and there are many open challenges to be tackled. In the following sections, we recommend several promising future research directions in the field of Collaborative Reinforcement Learning (CRL).

\paragraph{Combining Different Interactive Methods}
Develop more natural multi-feedback interactive methods by studying the advantages and disadvantages of different interactive methods. Single interactive methods would have higher requirements from humans and could be inefficient, whereas multi-modal interactive methods would lower interaction barriers and improve efficiency, providing users with a better interactive experience \cite{arumugam2019deep}. 

In the design patterns mentioned above, Combining Different Interactive Methods belongs to 'Augmentative Level collaboration'. It is the basis for the application of collaboration technology in real-life scenarios; it is also an important factor to improve user experience. Therefore, researchers could work on more advanced interaction modes based on this design concepts and applied to different scenarios.

\paragraph{User Modelling}
It is important to build generic user models to enable the system to accept user feedback robustly. Such models could be used to build human-AI collaboration applications that reduce human fatigue by detecting and predicting human behaviour \cite{arzate2020survey}, due to their ability of adapting to interaction channels and feedback types according to the user’s preferences. This would require empirical studies to find a way to map between user types and their preferred interaction channels and feedback types.

In the patterns mentioned above, user modelling is one of the most important issues of 'Integrative Level collaboration'. Only with accurate models could humans and AI agents communicate without barriers. The ability of predicting each other's behaviour could generate trust. Understanding the unexpected situations that may occur for each participant could establish a more flexible relationship and improve the entire system's robustness. Due to the rapid development of AI in recent years, there is still a lot of user modelling work that has not been carried out. Researchers could build AI and user models based on the perspective of 'Integrative Level collaboration': communication, trust, and resilience.

\paragraph{Lack of Human Collaborating Data and Evaluation Methods}
Human subjective is essential for improving CRL technology due to the inclusion of human involvement. However, there is limited research on human data collecting and evaluation \cite{arzate2021interactive}. This circumstance is caused by a variety of factors. For example, the expense of collecting human data is too high. Varied types of subjective human data need different techniques of collection, which is very challenging technically.

The typical approach to this problem is to conduct more experiments in order to collect sufficient data. Moreover, create evaluation techniques by merging several disciplines (e.g. psychology). However, there are certain innovative ways that need our attention. Strouse \textit{et al.} proposed a collaborating method without human data \cite{strouse2021collaborating}, which is called Fictitious Co-Play (FCP). In this work, They train the AI agent to be the best reaction by a population of self-play agents and their previous checkpoints taken during training. It might spark innovative thinking for training AI agents.

\paragraph{Safe Interactive RL}
Despite the empirical success of reinforcement learning algorithms, we have very little understanding of the way such 'black-box' models work. This means that the system cannot be responsible for their own decisions \cite{garcia2015comprehensive}. Therefore, how to establish a mechanism to protect human safety and avoid unknown discrimination becomes very important. Solving this problem is crucial for the use of interactive reinforcement learning in high-dimensional environments in the real world. 

In the patterns mentioned above, safe interactive RL is an important issue for the 'Integrative Level collaboration' and the 'Debative Level collaboration'. In terms of Integrative Level collaboration, how to ensure the safety of humans is one of the key factors for humans to trust AI. In terms of Debative level collaboration, when the decisions of humans and AI agents are inconsistent, how to protect the interests is not only an engineering issue, but also an ethical issue. This issue requires the joint efforts of multiple disciplines such as law, sociology, and ethics.

\paragraph{Explainable Collaborative Reinforcement Learning}
Since reinforcement learning needs to be trained via environmental feedback, different sensors or human feedback will lead to significantly different outcomes. Besides, explaining why the AI makes the specific decision is critical for humans to trust AI and allocate essential tasks to the AI agent. Therefore, developing the explainable collaborative reinforcement learning algorithm is required. 

Explainable collaborative RL approach aims to help people understand how the agent perceives the environment and how it makes decisions by improving the transparency of the model. Improving the transparency of the model not only allows humans to find better ways to train the agent, but also allows them to trust the agent to collaborate \cite{liu2019experience, krening2019effect, adadi2018peeking}.

\paragraph{High-dimensional Scenarios}
At present, both reinforcement learning and collaborative reinforcement learning have a relatively limited application scenario. The majority of experiments were conducted in virtual gaming environments. The agent is confronted with a modest dimensional environment, and there is a small number of variables. This complicates the implementation of CRL in high-latitude real-world scenarios. There is an algorithm that efficiently transfers human behaviour to the agent through an autoencoder \cite{arumugam2019deep}. Another possibility is to utilise crowd-sourcing to send human input \cite{isbell2001social}. However, this kind of issue requires more investigation.

\paragraph{Fast Evaluation of Human-Acts}
In the real scenario, many tasks need to be executed with rapid response, such as stock trading and autonomous driving. How to quickly evaluate the reaction of the other party to make the next decision is crucial. Some studies are currently focused on this area, and, in particular, exploring the methods of evaluating human behavior through visualisation models \cite{amir2018highlights, hadfield2017inverse,sorensen2016breeding, wilson2012bayesian}, but more forms of evaluation of human behaviour are needed.

\paragraph{Dynamic Mental Models}
The cooperative model requires a dynamic mental model from both humans and AI agents, as they constantly observe and learn about each other. It also needs to update strategies timely during the learning process \cite{frederiksen1999dynamic}. With the increase of collaboration experience, it would be useful to inject experience into the new learning process. At this stage, the implicit human prior knowledge needs to be gradually transformed into explicit experience guidelines for future tasks. Therefore, establishing a dynamic mental model may greatly promote the development of human-AI collaboration.

In the patterns mentioned above, dynamic mental models are critical to the three levels, making dynamic adjustments according to different collaboration levels, which has a great effect on reducing power consumption and improving efficiency \cite{schwalbe2020survey}. This is a huge challenge for researchers. A potential direction is to design general dynamic mental models according to the taxonomy we provide in this paper.

\section{Conclusions}
In this paper, we have presented a survey of Collaborative Reinforcement Learning (Collaborative RL, or CRL) to empower the research into human-AI interactions and cooperative designs. This analysis resulted in us proposing a \textit{new CRL classification method} (see Table 1), called \textit{CRL Design Trajectory Map} (see Figure \ref{map}) and a \textit{new CRL taxonomy} (see Figure \ref{category}) as a systematic modelling tool for selecting and improving new CRL designs. Researchers could use our Trajectory Map to design a CRL system from scratch or use parts of it according to their needs to refine their system. For example, researchers could select their desired system structure in Human-AI Collaborative Design Pattern, identify and satisfy the requirements of different components in Collaborative Levels \& Parties and Collaborative Capabilities, and select different design components in Algorithmic Models and Interactive Methods. This is a comprehensive design approach from top to bottom and from macro to micro. Summarising, through this survey, we provide researchers and practitioners with the tools to start improving and creating new designs for CRL methods.

\bibliographystyle{unsrt}  
\bibliography{references}

\end{document}